\documentclass[preprint2]{aastex61}

\usepackage{amsmath}
\newcounter{chemequation}
\renewcommand*\thechemequation{[R\arabic{chemequation}]}
\newenvironment{chemequation}{
\stepcounter{chemequation}
\begin{equation}}
{\tag*{\thechemequation}
\end{equation}}

\accepted{December 10, 2017}

\shorttitle{Atmospheric modeling of exo-Titans}
\shortauthors{Lora et al.}

\begin{document}

\title{Atmospheric circulation, chemistry, and infrared spectra of Titan-like exoplanets around different stellar types}

\correspondingauthor{Juan M. Lora}
\email{jlora@ucla.edu}

\author[0000-0001-9925-1050]{Juan M. Lora}
\affil{Department of Earth, Planetary, and Space Sciences, \\
University of California, Los Angeles, \\
595 Charles Young Drive, Los Angeles, CA 90095\\}

\author[0000-0003-3759-9080]{Tiffany Kataria}
\affiliation{Jet Propulsion Laboratory, \\
California Institute of Technology, \\
4800 Oak Grove Drive, Pasadena, CA 91109, USA\\}

\author[0000-0002-8518-9601]{Peter Gao}
\affiliation{Department of Astronomy, \\
University of California, Berkeley, \\
Berkeley, CA 94720, USA\\}



\begin{abstract}

With the discovery of ever smaller and colder exoplanets, terrestrial worlds with hazy atmospheres must be increasingly considered. Our Solar System's Titan is a prototypical hazy planet, whose atmosphere may be representative of a large number of planets in our Galaxy. As a step towards characterizing such worlds, we present simulations of exoplanets that resemble Titan, but orbit three different stellar hosts: G-, K-, and M-dwarf stars. We use general circulation and photochemistry models to explore the circulation and chemistry of these Titan-like planets under varying stellar spectra, in all cases assuming a Titan-like insolation. Due to the strong absorption of visible light by atmospheric haze, the redder radiation accompanying later stellar types produces more isothermal stratospheres, stronger meridional temperature gradients at mbar pressures, and deeper and stronger zonal winds. In all cases, the planets' atmospheres are strongly superrotating, but meridional circulation cells are weaker aloft under redder starlight. The photochemistry of hydrocarbon and nitrile species varies with stellar spectra, with variations in the FUV/NUV flux ratio playing an important role. Our results tentatively suggest that column haze production rates could be similar under all three hosts, implying that planets around many different stars could have similar characteristics to Titan's atmosphere. Lastly, we present theoretical emission spectra. Overall, our study indicates that, despite important and subtle differences, the circulation and chemistry of Titan-like exoplanets are relatively insensitive to differences in host star. These findings may be further probed with future space-based facilities, like WFIRST, LUVOIR, HabEx, and OST.

\end{abstract}




\section{Introduction} \label{sec:intro}

Over the past two decades, thousands of planets have been discovered around stars other than the Sun. Current exoplanet discoveries are increasingly moving toward smaller planets, and results from the NASA {\it Kepler} Space Telescope indicate that small planets ($\lesssim2~R_{\oplus}$) are the most common in our Galaxy \citep{fressin+2013}.  The discoveries of planets GJ 1132b \citep{berta-thompson+2015}, LHS 1140b \citep{dittmann+2017}, and the TRAPPIST-1 system \citep{gillon+2017} show that the characterization of small, rocky exoplanets is within reach, particularly around low-mass stars.  The NASA Transiting Exoplanet Survey Satellite (TESS, slated to launch in spring 2018) will identify small exoplanets orbiting bright stars with a range of stellar types.  Many of these targets will be favorable for follow-up characterization from the ground with future ELTs, as well as from space-based observatories like the Wide Field Infrared Survey Telescope (WFIRST, slated to launch in the mid-2020s)
and potential future large astrophysics mission facilities like the Large UV/Optical/Infrared Surveyor \citep[LUVOIR;][]{crooke+2016}, the Habitable Exoplanet Imaging Mission \citep[HabEx;][]{mennesson+2016}, and the Origins Space Telescope \citep[OST;][]{meixner+2016}. WFIRST, HabEx and LUVOIR would directly image planets in reflected light over UV--optical wavelengths, while OST would probe the thermal emission of transiting exoplanets in the near- to far-infrared (IR).

As small exoplanets continue to be discovered and characterized, we must rely on Solar System planets as analogues for understanding the characteristics of their atmospheres.  Previous studies, for example, have explored various aspects of the atmospheres and climates of putative terrestrial exoplanets---including their habitability, surface climate, and atmospheric circulation---with an emphasis on Earth-like conditions \citep{kaspi+showman2015,kopparapu+2013,merlis+shneider2010,shields+2013,shields+2016}. \citet{kaspi+showman2015} provide a comprehensive look at the atmospheric dynamics and climates of idealized terrestrial exoplanets (i.e., spherical aquaplanets with relatively thin atmospheres) over a wide range of physical and orbital parameters (planetary rotation rate, incident stellar flux, atmospheric mass, surface gravity, atmospheric optical thickness, and planetary radius), using an idealized general circulation model (GCM) with an Earth-like annual-mean climate as reference.  These results show that all of these parameters have considerable effects on the resulting latitudinal temperature differences, latitudinal heat transport, mean circulation, and hydrologic cycle of those atmospheres. 

While many investigations have based their simulations on Earth-like conditions, few investigations \citep[e.g.,][]{morley+2015} have explored the range of possible conditions in exoplanet atmospheres with complex photochemical hazes and at different equilibrium temperatures, such as those encountered on Titan.  In the Solar System, Titan's effective orbit around the Sun has a semi-major axis of about 9.5 AU, giving Titan an effective temperature of $\sim$80 K. Around lower mass (i.e., M dwarf) stars, the same effective temperature would be reached at considerably smaller semi-major axes ($\sim$1 AU). While Earth-like exoplanets would remain in the liquid-water habitable zone of M dwarf stars at orbits of 0.02--0.2 AU \citep{tarter+2007}, where issues of tidal locking and a heightened influence of stellar activity could pose problems for atmospheric stability and habitability \citep{khodachenko+2007}, exoplanets receiving Titan-like insolations around M dwarfs would be less affected by these obstacles.  Given the relative abundance of M to G stars in the Galaxy, Titan may therefore represent a more prototypical world with a stable hydrologic cycle than Earth \citep{lunine2010}. Titan-like exoplanets---cool terrestrial worlds with dense, hazy, and chemically complex atmospheres---would also be interesting subjects for studies of habitability, as Titan has been suggested to be an important target for astrobiology \citep{horst+2012,lunine2009,lunine2010,rahm+2016}.

Previous studies have applied one-dimensional radiative transfer models to explore Titan under different conditions. \citet{lorenz+1997} used a 1D model to calculate the surface temperature of Titan as a function of stellar insolation, as a proxy for the Sun's evolution. They found that during the red giant phase, solar UV flux is low enough for haze production to decrease, such that more sunlight could reach Titan's surface, potentially allowing life to form.  \citet{gilliam+mckay_2011} used a similar 1D model to calculate the distance at which a putative planet would maintain a Titan-like surface temperature when orbiting two different types of M dwarf stars, varying the atmospheric haze production over four orders of magnitude.

More recently, the studies of \citet{robinson+2014} and \citet{checlair+2016} illustrated the potential for future observational characterization of Titan-like exoplanets.  \citet{robinson+2014} used Cassini/VIMS occultation observations to understand what Titan would look like in transit (i.e, passing in front of the Sun along the line of sight of some faraway observer).  They found that the continuum level and slope of the transit spectrum are largely set by Titan's haze. \citet{checlair+2016} calculated geometric albedo and effective transit spectra as a function of haze production rate, and found that haze production leaves a clear signature in the slopes of albedo spectra, particularly between the UV and visible wavelengths, which are more sensitive to the production rate than transit spectra. Spectra therefore represent a useful tool for future characterization of cool, hazy terrestrial exoplanets. But these spectra are also dependent on the structure, composition, and dynamics of these atmospheres; in the thermal infrared, for instance, observable spectra are strongly dependent on the atmospheric temperature structure.

In this paper, we explore the circulation, photochemistry, and resulting observable emission spectra of the atmospheres of Titan-like exoplanets around different stellar types. To this end, we adapt and apply an existing Titan GCM and photochemical model to explore the thermal, dynamical, and chemical structure of Titan-like exoplanets as a function of their host stellar type.  In Section~\ref{sec:methods}, we present the models we use for our simulations, TAM and KINETICS.  In Section~\ref{sec:results}, we describe the results from both sets of models. We present a discussion of our results and conclude in Section~\ref{sec:conclusion}.

\section{Methods} \label{sec:methods}

\subsection{TAM}
The GCM simulations presented here are run with the Titan Atmospheric Model (TAM), a climate model with physics parameterizations for planetary applications that has been validated with observations from the NASA-ESA Cassini-Huygens mission \citep{lora+2015}. Briefly, TAM employs the Geophysical Fluid Dynamics Laboratory's (GFDL) spectral dynamical core \citep{gordon_stern1982} to solve the primitive equations in vorticity-divergence form, and includes parameterizations for radiative transfer, surface, and boundary layer processes. In our simulations, TAM is run in its L50 configuration (50 unevenly-spaced vertical layers extending from the surface around 1500 mbar to the lower mesosphere at about 3 $\mu$bar), and at T21 horizontal resolution (approximately 5.6$^{\circ}$ resolution) which minimizes computational costs but is sufficient to attain superrotation \citep{lora+2015}.

The radiative transfer is computed using fully nongray, multiple-scattering two-stream approximations \citep{toon+1989}. Opacities due to molecular absorbers are treated using correlated {\it k} coefficients computed from HITRAN line intensities \citep{rothman+2009}, and include the effects of methane, acetylene, ethylene, ethane, and hydrogen cyanide. Opacity due to collisionally-induced absorption from methane, molecular nitrogen, and molecular hydrogren pairs is considered \citep{richard+2011}. Extinction due to Titan's characteristic haze is treated with the model of \citet{tomasko+2008}, and Rayleigh scattering is also included. 

TAM has been used to study the methane-based hydrologic cycle in Titan's troposphere \citep{lora+2014,lora_mitchell2015,lora_adamkovics2017}, but in this work we ignore moist thermodynamics and the influence of regionally-varying surface liquids. We assume in all cases that methane is resupplied to the atmosphere to prevent its eventual loss through escape (see Section~\ref{s:CH4}), such that the middle troposphere is always close to saturation, as on Titan. Atmospheric methane is considered for the purposes of radiative transfer, with profiles determined as described below. The model does not include the effects of condensation clouds, which are infrequent and confined to the lower atmosphere.

\subsection{KINETICS}
We investigate the photochemistry of Titan-like exoplanets using the Caltech/JPL 1--D photochemical and transport model KINETICS \citep{allen+1981}. KINETICS solves the 1--D continuity equation for each chemical species, taking into account chemical production and loss rates, and transport in a 1--D column due to molecular and eddy diffusion. KINETICS has been used to model Titan photochemistry since its inception \citep{yung+1984}, and more recently it has been validated using Cassini observations \citep{li+2014,li+2015} that constrained Titan's atmospheric eddy diffusivity and the chemical pathways controlling the abundances of $\mathrm{C_2}$ and $\mathrm{C_3}$ hydrocarbons. 

The present Titan KINETICS model stems from \citet{li+2015}, which includes 379 reactions and 65 chemical species incorporating both hydrocarbon and nitrile chemistry up to $\mathrm{C_6}$ species and $\mathrm{C_{4}N_{2}}$. The altitude range of the model extends from 45 to 1447 km, bypassing condensation processes in the troposphere below 45 km while capturing the photolytically active region in the upper atmosphere. We use the eddy diffusion coefficient of \citet{li+2014} for all cases and iterations (see below). We maintain a zero-flux upper boundary condition except for H and $\mathrm{H_2}$, which are allowed to escape to space assuming escape is diffusion limited. The lower boundary is assumed to possess a constant concentration gradient for almost all species allowing for free exchange with the atmosphere below. Exceptions to this include methane, which has a fixed mixing ratio of 1.4\%, and H and $\mathrm{H_2}$, which have zero flux boundary conditions.

\subsection{Model Setup}
In order to analyze the effects of circulation and chemistry on Titan-like exoplanets orbiting different stellar types, we first run TAM simulations under the influence of insolation from three spectrally different stars: a G dwarf (the Sun, G2V, which acts as our control case), a typical K dwarf (HD85512, K6V), and a typical M-dwarf (GJ436, M3.5V). The K and M dwarf spectra are from \citet{loyd+2016}.

In each case, we assume the planet is in a perfectly circular orbit with zero obliquity, such that average top-of-atmosphere insolation is constant and symmetric about the equator. This produces a climate under perpetual ``equinox'' conditions, in which seasonal variations do not occur. In order to isolate in particular the effects of the stellar spectra, we assume a Titan-like incident stellar flux (15 ${\rm W~m^{-2}}$) in all cases. Equivalently, this corresponds to planets at 9.5, 3.75, and 1.55 AU from the G, K, and M host stars, respectively.

For all cases, the planetary parameters---planetary radius, planetary rotation, surface gravity, background gas---are held constant at values corresponding to Titan \citep{lora+2015}, in order to limit the complexity of parameter space so that we can unambiguously determine the atmospheric sensitivity to stellar host forcing conditions. Diurnal cycles are included in the computation of insolation, but otherwise all forcing is axisymmetric (and symmetric between northern and southern hemispheres). Diurnal cycles are not included in photochemical calculations. Inhomogeneities in surface properties, including any topographic variations, are also ignored for simplicity.

We perform an asynchronous coupling between dynamics and chemistry by first running TAM with default gaseous species profiles \citep{vinatier+2007}, then running KINETICS with an updated temperature structure from the GCM and the appropriate stellar spectrum to compute new species profiles, re-running TAM with the new profiles, and iterating between the two models. The temperature structure at altitudes above the model domain of TAM is held fixed in KINETICS for all iterations. Dynamical effects on the horizontal distribution of the minor species are ignored, so the profiles represent global averages. Results from KINETICS are primarily dependent on the stellar spectrum, and the radiative effects in TAM are relatively insensitive to small changes in absorber profiles, so convergence between the two models is achieved quickly. In all cases, TAM is run from rest (zero winds) for approximately 2,000 Earth years in total in order to achieve a steady state circulation in which the atmospheric angular momentum reaches a long-term steady state. 

\begin{figure}
   \centering
    \includegraphics[]{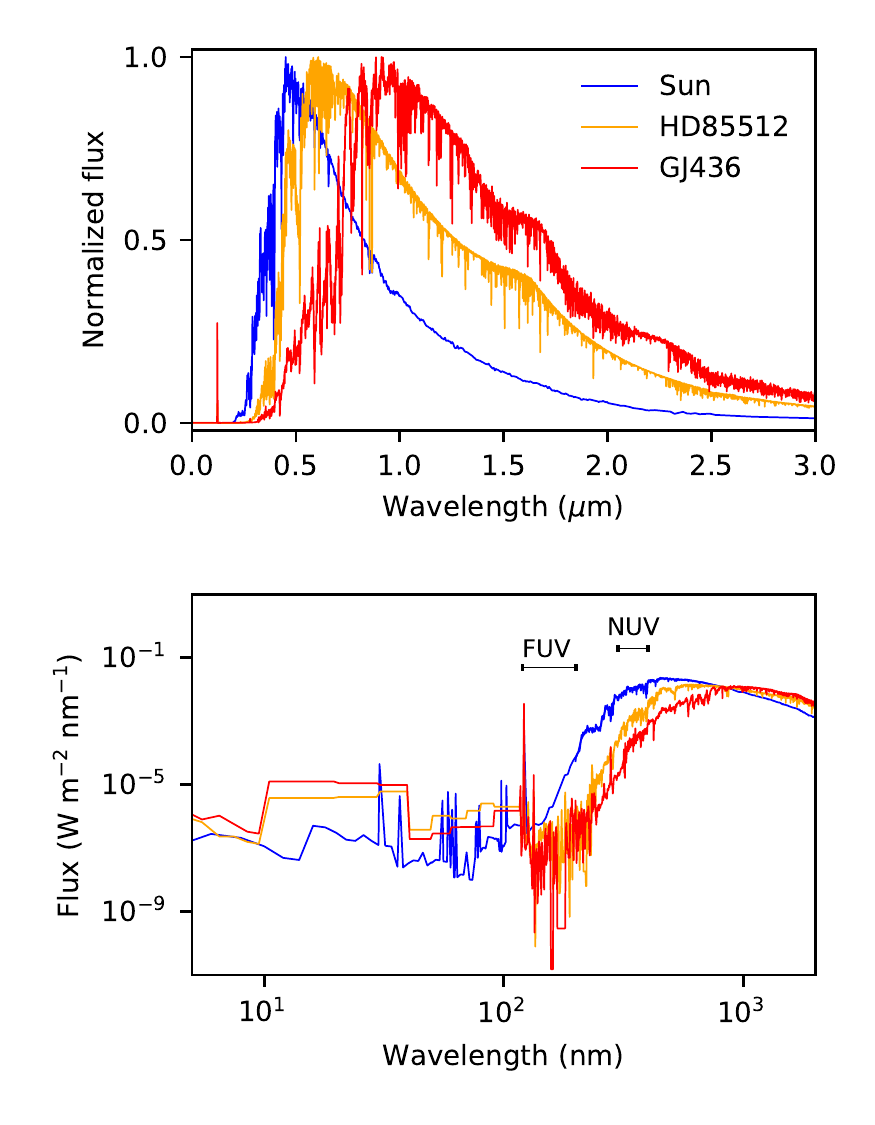}
   \caption{Stellar host spectra. (Top) UV to near-IR stellar spectra for a G dwarf (the Sun; blue curve), a K dwarf (HD85512; orange curve), and an M dwarf (GJ436; red curve), normalized to their maximum fluxes. (Bottom) The same stellar spectra, formatted to highlight differences at short wavelengths, and showing the fluxes at the various planets' distances (see text).}\label{stellar_spectra}
\end{figure}

Comparing the stellar spectra in Figure \ref{stellar_spectra}, the peak in stellar radiation moves to longer wavelengths between K and G, and M and K stars.  This wavelength dependence affects both the resultant circulation, through its impacts on the location of absorption of insolation, and chemistry. At UV wavelengths, the FUV/NUV ratio of the M dwarf is greater than that of the K dwarf, which is in turn greater than that of the G dwarf \citep{france+2016}, indicative of high stellar activity for the cooler stars. In particular, the Lyman-$\alpha$ flux at the location of the Titan-like exoplanet is comparable between that of the K and M dwarfs, both of which are an order of magnitude greater than that of the G dwarf. These variations can have a large impact on photochemistry due to the wavelength dependence of photolysis cross sections. For example, while methane is photolyzed mostly in the FUV, higher order hydrocarbons can be photolyzed by both FUV and NUV photons \citep{hebrard+2013,lavvas+2008a,wilson+atreya2004}, setting up autocatalytic pathways that lead to higher rates of methane loss \citep{yung+1984}. Disruption of this chemical scheme due to variations in the FUV/NUV ratio could lead to significant differences in abundances. 

The haze distribution is held constant between all TAM simulations, since haze production rates and their dependence on stellar type are as yet highly uncertain \citep{gilliam+mckay_2011,checlair+2016}, and our focus here is on the resulting profiles of low-order hydrocarbons, as well as the global circulation.

\section{Results} \label{sec:results}

\subsection{Circulation and Temperature Structure} 

\begin{figure*}
   \centering
    \includegraphics[]{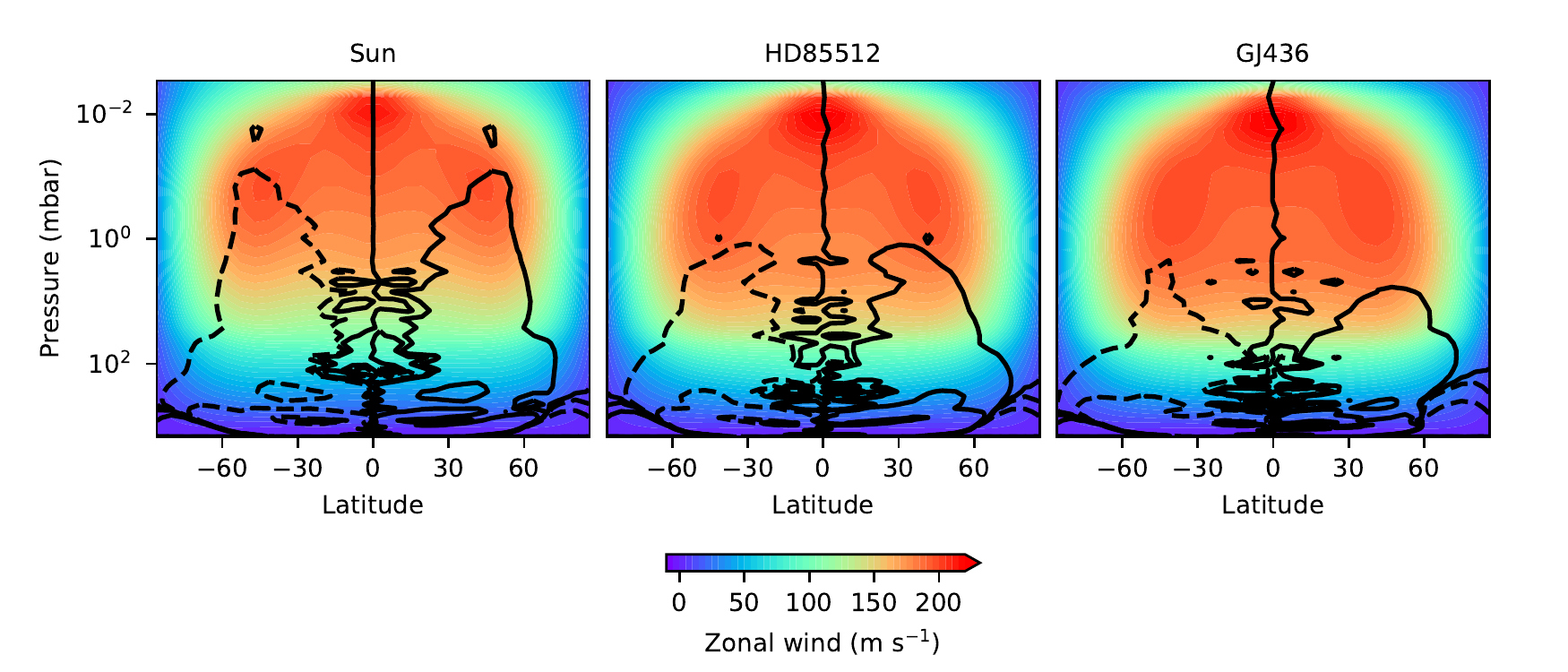}
   \caption{Atmospheric circulations. Zonal-mean zonal winds (colors) and meridional overturning mass streamfunctions (black contours) for planets around G (left), K (middle), and M (right) stars. Contours show the 0, 5$\times$10$^7$, and 5$\times$10$^8$\,kg\,s$^{-1}$ streamlines, with dashed curves indicating counter-clockwise motion.}\label{circulation}
\end{figure*}

A principal characteristic of Titan's atmospheric circulation is the presence of strong superrotation, where the atmosphere has more angular momentum than the solid body at the equator \citep{mitchell_vallis2010}; TAM has successfully reproduced this superrotation \citep{lora+2015} in comparison to observations of Titan \citep{achterberg+2008}. Fig.~\ref{circulation} shows the zonally averaged zonal wind in each of our simulations, demonstrating that the atmospheric superrotation is a robust feature regardless of the host stellar spectrum. Interestingly, peak zonal winds---which presumably occur over the equator as a consequence of angular momentum convergence by waves \citep{newman+2011}---are weakest for the case of a solar-type stellar host, and increase in strength with redder host stellar spectra. Though the peak winds occur at the same pressure levels (approximately 10 $\mu$bar in all three cases), strong winds extend to deeper levels for the K and M host stars. This is particularly true for the mid-latitude jets. 

The zonally averaged meridional circulation for the three simulations is also shown in Fig.~\ref{circulation} (black contours). In all cases, single circulation cells dominate each hemisphere, as is the case for Titan at equinox \citep{lebonnois+2012,lora+2015,newman+2011}, as a result of the slow planetary rotation \citep[see][]{kaspi+showman2015}. At pressures above approximately 1 mbar, these circulation cells are thermally direct. Poleward transport of angular momentum occurs in the upper branches of these cells, which prevail at increasingly deep levels for G, K, and M host stars. Thus, the depth of the meridional circulation is related to the depth of the mid-latitude jets, impacting higher pressures for redder host star spectra.

Shallow, thermally indirect meridional circulation cells also occur in the polar lower tropospheres of all simulations, extending from the pole to roughly 60$^{\circ}$ latitudes. These cells appear as the result of time averaging of regions where shallow baroclinic eddies are present \citep{lebonnois+2012,lora_mitchell2015}, as in Earth's mid-latitudes. The exact extent of these cells varies and depends on the averaging interval, but we find no dependence of these features on the spectrum of the host star.

\begin{figure*}
   \centering
    \includegraphics[]{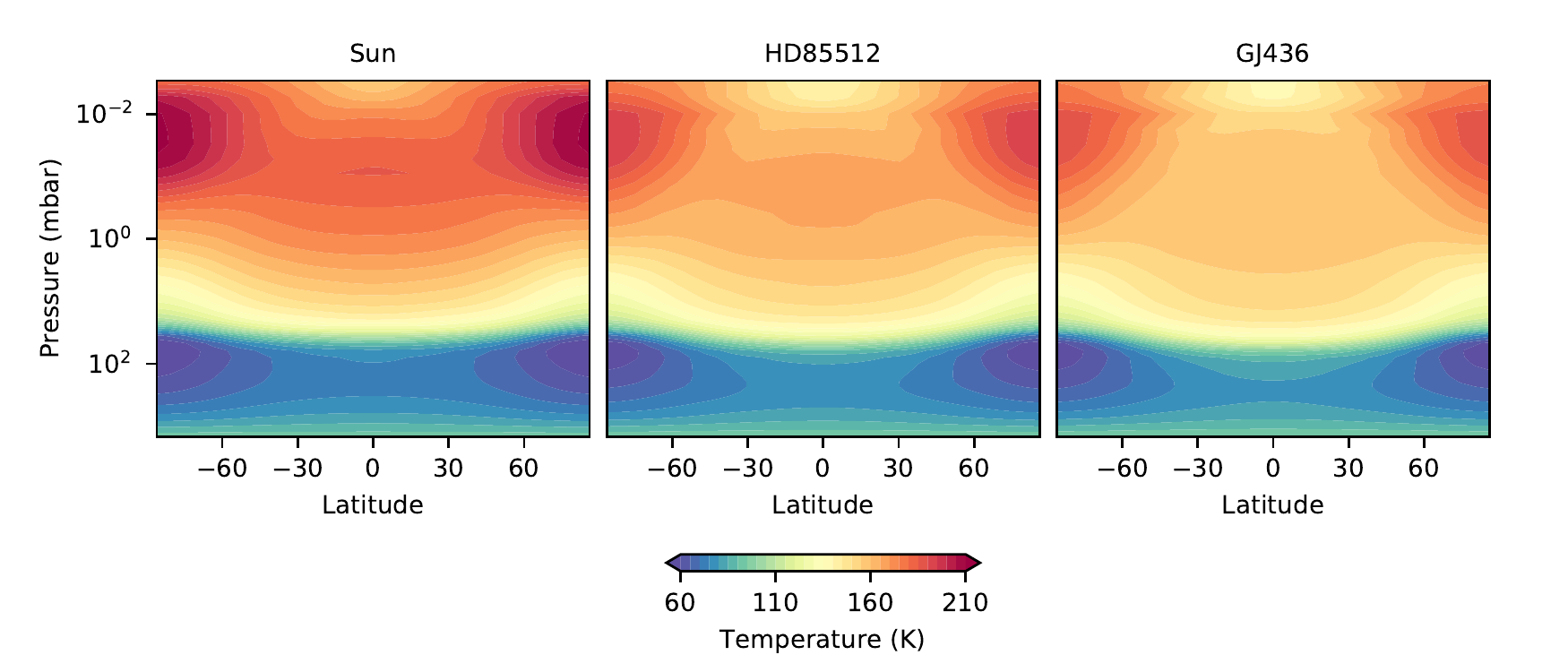}
   \caption{Atmospheric temperatures. Zonal-mean temperatures for planets around G (left), K (middle), and M (right) stars.}\label{temps}
\end{figure*}

Zonally-averaged temperatures for the three TAM simulations are shown in Fig.~\ref{temps}. The atmospheric structure is similar in all cases, with a defined troposphere and stratosphere, as well as the lowermost section of a mesosphere, in general agreement with simulations of Titan \citep{lora+2015}. At polar latitudes, the regions around 100 mbar sustain the coldest temperatures, which are approximately 10 K colder than at lower latitudes. Directly above these regions, roughly at 0.1 to 0.01 mbar, polar stratospheric temperatures are the warmest in the atmosphere at about 200 K. These warm regions are comparable to those observed in Titan's winter hemisphere \citep{achterberg+2008}, and form as the result of adiabatic heating of descending air driven by the meridional circulation \citep{lora+2015}. In these simulations, the high degree of symmetry between the hemispheres is due to the absence of seasonality and the resulting continuous double-celled structure of the circulation.

\begin{figure*}
   \centering
    \includegraphics[]{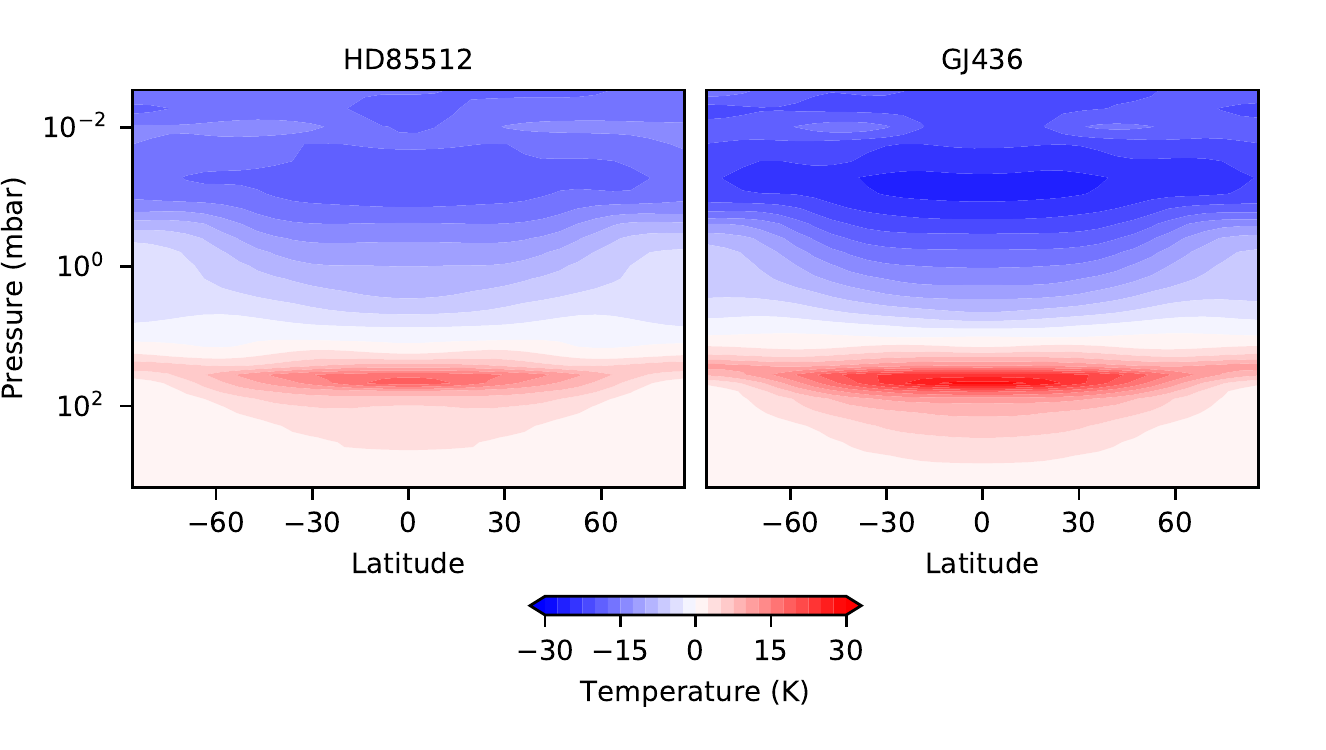}
   \caption{Atmospheric temperature differences. Differences in zonal-mean temperatures between planets around K and G stars (left), and M and G (right) stars.}\label{temp_diffs}
\end{figure*}

Temperature differences between the simulations occur at all pressure levels of the atmosphere (Fig.~\ref{temp_diffs}). An immediately apparent difference is that the stratospheres are relatively colder for simulations under redder starlight, while their respective tropospheres are relatively warmer than the case forced with solar radiation. These differences result from the wavelength-dependent absorption by haze in the atmosphere, which is strongest at short wavelengths and decreases toward near-IR wavelengths, therefore allowing more radiation from the lower-mass stars to penetrate to deeper levels \citep{checlair+2016,lorenz+1997}. In this regard, the dependence of the atmospheric thermal structure on host type is itself strongly dependent on the presence of haze.

The simulated high-latitude ($>$60$^{\circ}$) cold and hot regions vary in their relative magnitude with the host star spectra. The warmest regions occur in the stratosphere of the G host star case. Additionally, the G host star produces the strongest latitudinal temperature gradients in the uppermost stratosphere (while the weakest gradients there happen when irradiated by the M host star). 

In the troposphere, the coldest regions also occur in the atmosphere irradiated by the G host star. But with this host star, the upper troposphere and much of the stratosphere sustain meridional temperature contrasts that are weak, while the M host star causes the strongest gradients. Thus, the deposition of stellar radiation at deeper pressures in the case of K and M stars results in vertical wind shear that is strongest in these regions, in turn producing the deeper and stronger mid-latitude jets.

\begin{figure}
   \centering
    \includegraphics[]{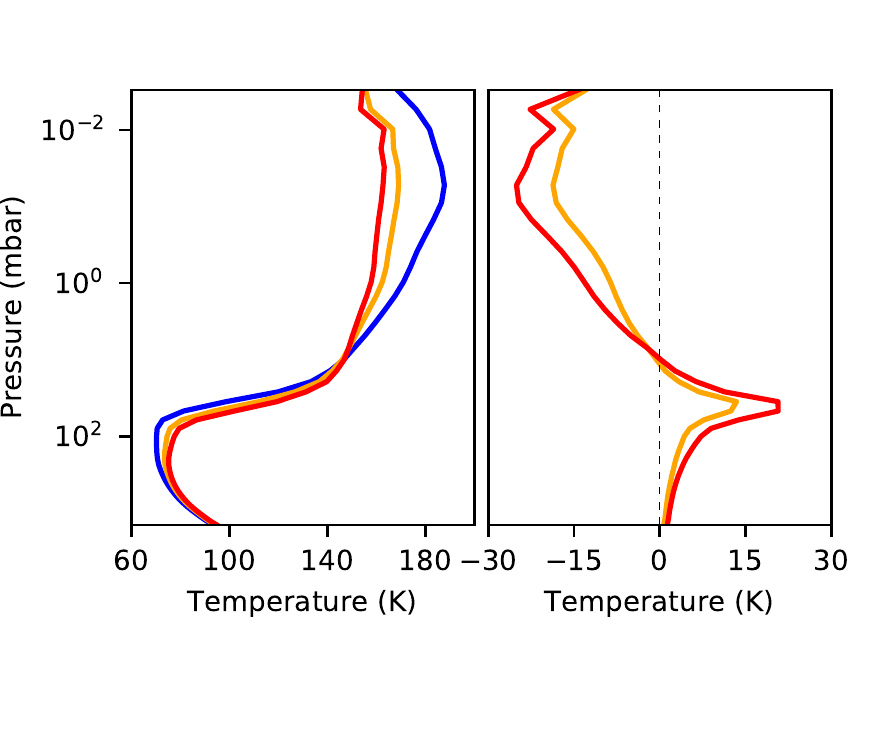}
   \caption{Atmospheric temperature profiles. Left: Global-mean temperature profiles for planets around G (blue curve), K (orange curve), and M (red curve) stars. Right: Difference in global-mean temperatures between planets around K and G stars (orange curve), and M and G (red curve) stars.}\label{temp_profiles}
\end{figure}

A comparison of global average temperature profiles (Fig.~\ref{temp_profiles}) further illustrates the progression from warm to cold stratospheres and cold to warm tropospheres between simulations forced by G, K, and M stellar spectra. The latter two simulations additionally produce stratospheres that are considerably more isothermal than that forced by the hottest star. Furthermore, these profiles show that the average pressure level of the stratopause is the same between simulations, but the largest temperature gradients in the lower stratospheres for planets orbiting the redder host stars occur at deeper pressures than the solar case.  Similarly, the minimum temperatures in the cases forced by redder stars occur at higher pressures.  The temperatures of the lower troposphere and surface in all three cases are approximately the same, suggesting that differences in the partitioning of energy absorption between the three cases occur primarily in the atmosphere.

\begin{figure*}
   \centering
    \includegraphics[]{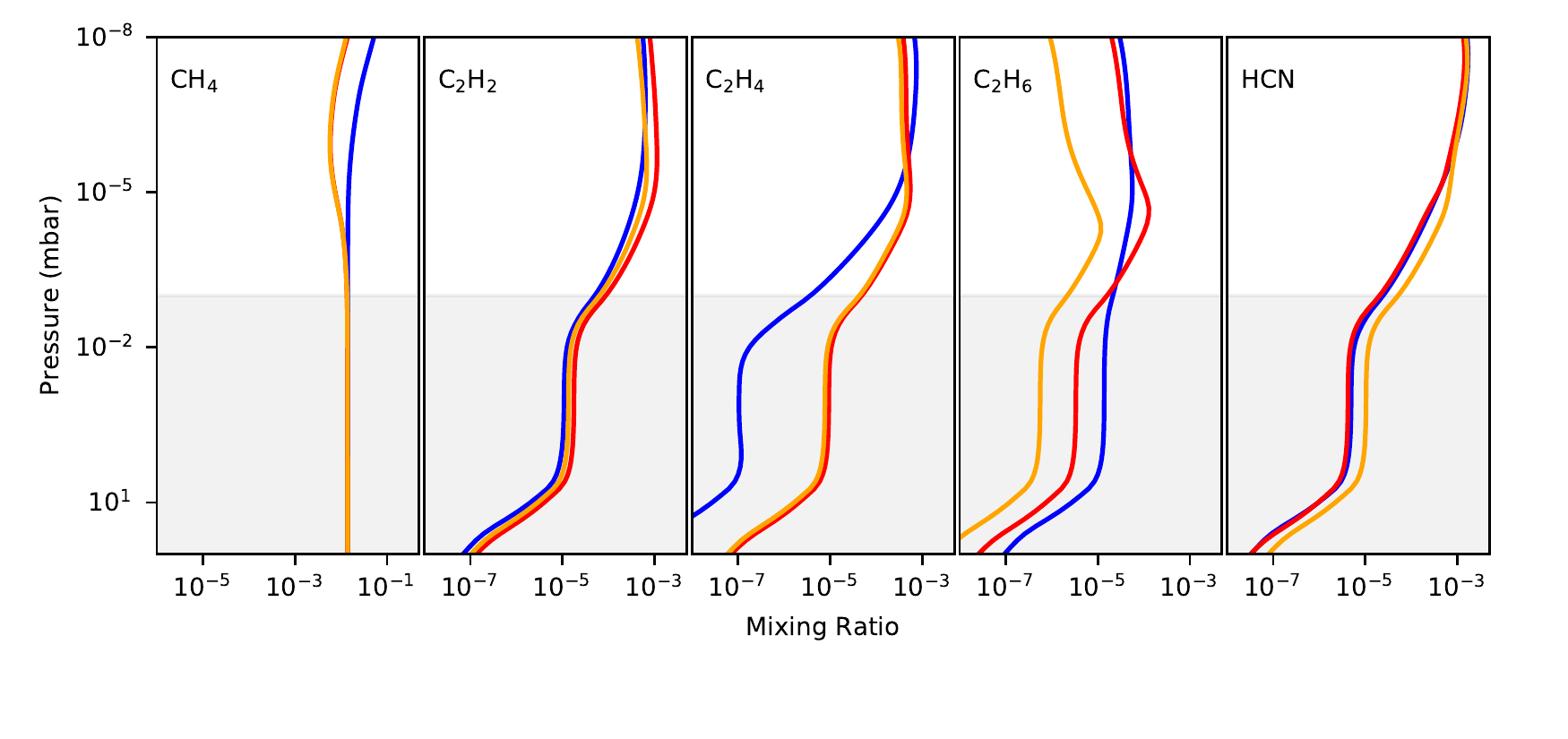}
   \caption{Trace gas profiles. Profiles of hydrocarbons and HCN in the atmospheres of planets around G (blue curves), K (orange curves), and M (red curves) stars. Lightly shaded regions show pressure ranges captured in the GCM.}\label{tracegasprofiles}
\end{figure*}

\begin{figure*}
   \centering
    \includegraphics[]{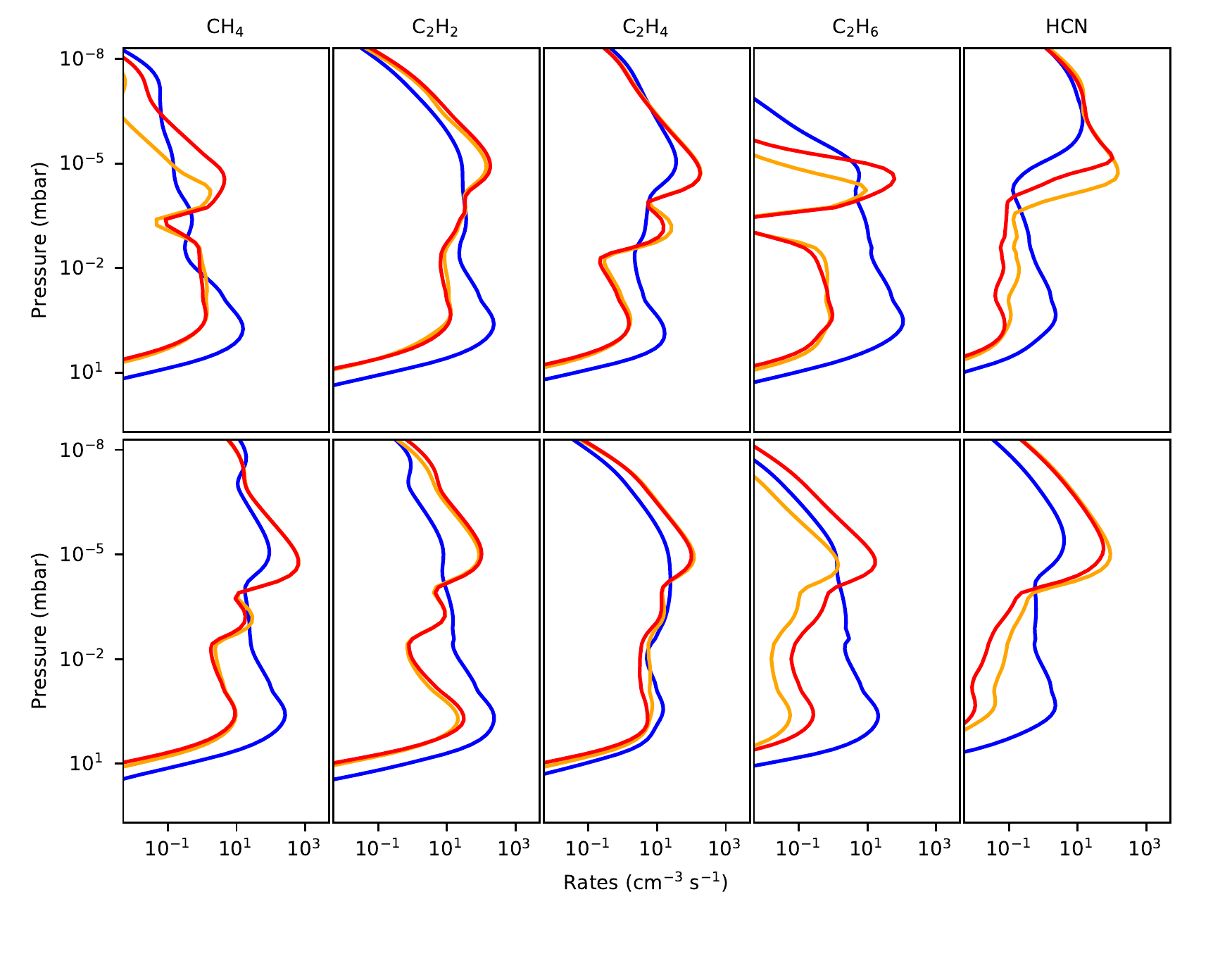}
   \caption{Trace gas production/loss rates. Production (top) and loss (bottom) rate profiles of photochemical trace gases for planets around G (blue curves), K (orange curves), and M (red curves) stars.}\label{prodlossrates}
\end{figure*}

\subsection{Photochemistry}
The mixing ratio profiles of the major hydrocarbon species and HCN are altered in different ways by the different stellar spectra (Fig.~\ref{tracegasprofiles}). While acetylene and HCN are changed by only a factor of a few, order of magnitude changes are seen for methane, ethylene, and ethane. This is a direct result of the different chemical pathways that lead to their production and destruction, and how they are affected by variations in the FUV/NUV ratio. 

\subsubsection{Methane ($CH_4$)}\label{s:CH4}

For all three stellar types the production rate of methane is dwarfed by the loss rate (Fig.~\ref{prodlossrates}), leading to net methane loss from the atmosphere through hydrogen escape. For the K and M dwarf cases methane is depleted above $10^{-4}$ mbar as compared to the G dwarf case, which results from photolysis by the 10 times higher Lyman-$\alpha$ flux around the cooler stars. The increased photolysis leads to higher production rates of reactive species such as CH, $\mathrm{CH_2}$, and $\mathrm{CH_3}$, which contribute to $\mathrm{C_2}$ hydrocarbon production and loss. Conversely, methane loss due to reaction with $\mathrm{C_2H}$ at 1 mbar, a major part of the methane autocatalytic scheme for the G dwarf case, is unimportant for the K and M dwarf cases. This arises due to reduced $\mathrm{C_2H}$ production below $10^{-4}$ mbar from acetylene photolysis, the rate of which is lowered for  cooler stars due to their lower NUV fluxes \citep{huebner+1992,wilson+atreya2004}. 

\subsubsection{Acetylene ($C_2H_2$)}

Acetylene mixing ratios are not greatly affected by changing stellar types. For the G dwarf case acetylene is mostly produced at 1 mbar from the reaction 

\begin{chemequation}
  \text{C}_2\text{H} + \text{CH}_4  \longrightarrow \text{C}_2\text{H}_2 + \text{CH}_3
  \label{ch:c2h2prod}
\end{chemequation}

\noindent and lost at the same pressure level from the related photolysis reaction 

\begin{chemequation}
  \text{C}_2\text{H}_2 + h\nu   \longrightarrow \text{C}_2\text{H} + \text{H}
  \label{ch:c2h2loss}
\end{chemequation}

\noindent These two reactions lead to the autocatalytic destruction of methane. However, as previously mentioned, this reaction is not relevant for the K and M dwarf cases. Instead, for these stellar hosts acetylene is produced primarily above $10^{-4}$ mbar through

\begin{chemequation}
  \text{H} + \text{C}_2\text{H}_3    \longrightarrow \text{C}_2\text{H}_2 + \text{H}_2
  \label{ch:c2h2prodkm1}
\end{chemequation}

\begin{chemequation}
  \text{CH}_3 + \text{C}_3\text{H}_2    \longrightarrow \text{C}_2\text{H}_2 + \text{C}_2\text{H}_3
  \label{ch:c2h2prodkm2}
\end{chemequation}

\noindent and lost through

\begin{chemequation}
  \text{CH} + \text{C}_2\text{H}_2    \longrightarrow \text{C}_3\text{H}_2 + \text{H}
  \label{ch:c2h2losskm}
\end{chemequation}

\noindent where both CH and $\mathrm{CH_3}$ are produced from methane photolysis by Lyman-$\alpha$. Reaction \ref{ch:c2h2losskm} is also an important loss mechanism for acetylene above $10^{-4}$ mbar in the G dwarf case, but its rate is an order of magnitude lower and it is dwarfed by \ref{ch:c2h2loss}. Thus, while the acetylene abundance is maintained through $\mathrm{C_2H}$ cycling in the G dwarf case, similar acetylene abundances are maintained through $\mathrm{C_2H_3}$ and $\mathrm{C_3H_2}$ cycling in the K and M dwarf cases stemming from the increased production of CH and $\mathrm{CH_3}$ from higher Lyman-$\alpha$ fluxes. 

\subsubsection{Ethylene ($C_2H_4$)}

Ethylene abundances are two orders of magnitudes higher for the K and M dwarf cases than for the G dwarf case. For all three cases the ethylene production is dominated by

\begin{chemequation}
  \text{CH} + \text{CH}_4
  \longrightarrow \text{C}_2\text{H}_4 + 	     \text{H}
  \label{ch:c2h4prod}
\end{chemequation}

\noindent where the higher production of CH in the K and M dwarf cases afforded by their higher Lyman-$\alpha$ fluxes lead to higher production rates of ethylene. In the G dwarf case, ethylene is primarily lost due to photolysis, forming acetylene and 2 H atoms, while for the K and M dwarf cases 

\begin{chemequation}
  \text{CN} + \text{C}_2\text{H}_4
  \longrightarrow \text{C}_2\text{H}_3\text{CN} + 	     \text{H}
  \label{ch:c2h4lossmk}
\end{chemequation}

\noindent also plays a major role. However, while the photolysis rate of ethylene in the K and M dwarf cases are higher than that in the G dwarf case, this is due to higher ethylene abundances rather than higher rate coefficients. Indeed, the rate coefficient of ethylene is higher in the NUV \citep{huebner+1992}, where K and M dwarfs lack photons, than much of the FUV. Therefore, the increased production of ethylene in the K and M dwarf cases caused by increased CH production from methane photolysis, coupled with the decrease in its photolysis rate coefficient, leads to a large increase in ethylene abundance in the cooler star cases compared to the G dwarf case. 

\subsubsection{Ethane ($C_2H_6$)}

The ethane abundance is different across all three stellar types, though with similar profile shapes in the K and M dwarf cases. For all three cases ethane is produced by 

\begin{chemequation}
  2\text{CH}_3 + \text{M}
  \longrightarrow \text{C}_2\text{H}_6 + 	     \text{M}
  \label{ch:c2h6prod}
\end{chemequation}

\noindent However, the production rate profile is different between the G dwarf case and the cooler star cases. In the former, the production peaks at 1 mbar, and it is caused by high rates of $\mathrm{CH_3}$ production from \ref{ch:c2h2prod}, which again plays a much more minor role in the K and M dwarf cases due to low NUV fluxes. Meanwhile, the production rate peak above $10^{-4}$ mbar in the K and M dwarf cases are due to increased production of $\mathrm{CH_3}$ from methane photolysis. The difference in ethane production rate (and thus ethane abundances) between the K and M dwarf cases is due to higher fluxes of photons with wavelengths between 50 and 100 nm from the K dwarf, which causes higher rates of nitrogen atom production from $\mathrm{N_2}$ photolysis. The excess N then leads to higher rates of $\mathrm{CH_3}$ loss via

\begin{chemequation}
  \text{N} + \text{CH}_3 
  \longrightarrow \text{HCN} + \text{H}_2
  \label{ch:ch3loss}
\end{chemequation}

\noindent and results in fewer $\mathrm{CH_3}$ to participate in ethane formation. 

Ethane is destroyed in the G dwarf case by 

\begin{chemequation}
  \text{C}_2\text{H} + \text{C}_2\text{H}_6
  \longrightarrow \text{C}_2\text{H}_2 + \text{C}_2\text{H}_5
  \label{ch:c2h6loss1}
\end{chemequation}

\begin{chemequation}
  \text{C}_3\text{N} + \text{C}_2\text{H}_6
  \longrightarrow \text{HC}_3\text{N} + \text{C}_2\text{H}_5
  \label{ch:c2h6loss2}
\end{chemequation}

\noindent at 1 mbar. Both of these reactions are slowed in the cooler star cases due to reduced NUV. Instead, in those cases ethane is destroyed mainly above $10^{-4}$ mbar by 

\begin{chemequation}
  \text{CH} + \text{C}_2\text{H}_6
  \longrightarrow \text{C}_3\text{H}_6 + \text{H}
  \label{ch:c2h6losskm1}
\end{chemequation}

\begin{chemequation}
  \text{CN} + \text{C}_2\text{H}_6
  \longrightarrow \text{HCN} + \text{C}_2\text{H}_5
  \label{ch:c2h6losskm2}
\end{chemequation}

\noindent with CH and CN arising from increased methane and HCN photolysis due to Lyman-$\alpha$, respectively.  

\subsubsection{Hydrogen Cyanide (HCN)}

In all three stellar cases HCN is produced mostly above $10^{-4}$ mbar through \ref{ch:ch3loss} and destroyed by photolysis to form H and CN by mainly Lyman-$\alpha$. The small peak in the HCN loss rate at 1 mbar in the G dwarf case is due to slightly higher HCN photolysis rates there caused by the higher NUV flux. The similarity of the HCN abundances between the three different stellar cases is due to both HCN formation and loss being tied to the Lyman-$\alpha$ flux. In other words, as $\mathrm{CH_3}$ is the limiting factor in \ref{ch:ch3loss} and is formed from Lyman-$\alpha$ photolysis of methane, increasing Lyman-$\alpha$ results in higher HCN formation rates, but also higher HCN photolysis rates. HCN abundances are slightly higher for the K dwarf case due to higher N abundances from increased flux of photons with wavelength $<$100 nm.

\subsubsection{Haze}

\citet{lavvas+2008a,lavvas+2008b} parameterized the haze formation rate in Titan's atmosphere as the combined rates of reactions where one of the products is a hydrocarbon with ten or more carbon atoms. Higher order nitrile species are also included, but they form a small fraction of the total haze formation rate and so will be ignored here. 

\begin{figure}
   \centering
    \includegraphics[]{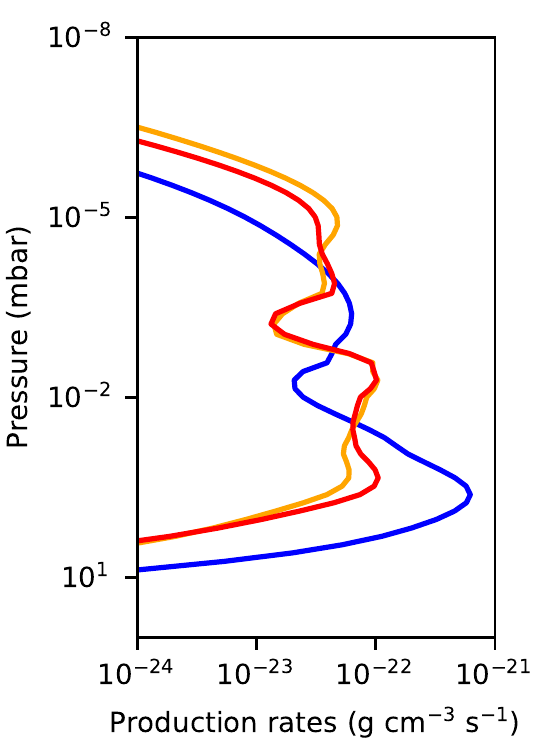}
   \caption{Production rate profiles of photochemical haze. Profiles for planets around G (blue curves), K (orange curves), and M (red curves) stars, assuming the haze production mechanism outlined in \citet{lavvas+2008a}.}\label{hazerates}
\end{figure}

Using this proxy, we compare the estimated haze formation rate profiles between the three stellar cases in Fig.~\ref{hazerates}, where we have converted the reaction rates in units of g cm$^{-3}$ s$^{-1}$ to a mass production rate of $\mathrm{C_{10}H_{22}}$. The K and M dwarf profiles are very similar and are located at a slightly higher part of the atmosphere (about an order of magnitude lower in pressure). In contrast, the G dwarf profile sits lower in the atmosphere and features a large peak at 1 mbar, which is consistent with the maxima in $\mathrm{C_2}$ hydrocarbon production/loss due to reactions powered by NUV fluxes (Fig.~\ref{prodlossrates}). This suggests that the NUV/FUV ratio of the host star affects the location and mass loading of the haze. However, while the column integrated haze production rate for the G dwarf is $0.8\times 10^{-14}$ g cm$^{-3}$ s$^{-1}$, consistent with the haze formation rate estimated by \citet{mckay+2001} of $0.5-2\times 10^{-14}$ g cm$^{-3}$ s$^{-1}$, the same quantity for the K and M dwarfs are only slightly lower, at $0.3\times 10^{-14}$ g cm$^{-3}$ s$^{-1}$. Given the unknowns in the haze formation pathway, it is difficult to conclude that the haze formation rate is significantly different between the three stellar cases. Therefore, our prior assumption of similar haze distributions across the three stellar types may be valid. 

\subsection{Atmospheric Spectra}

\begin{figure}
   \centering
    \includegraphics[]{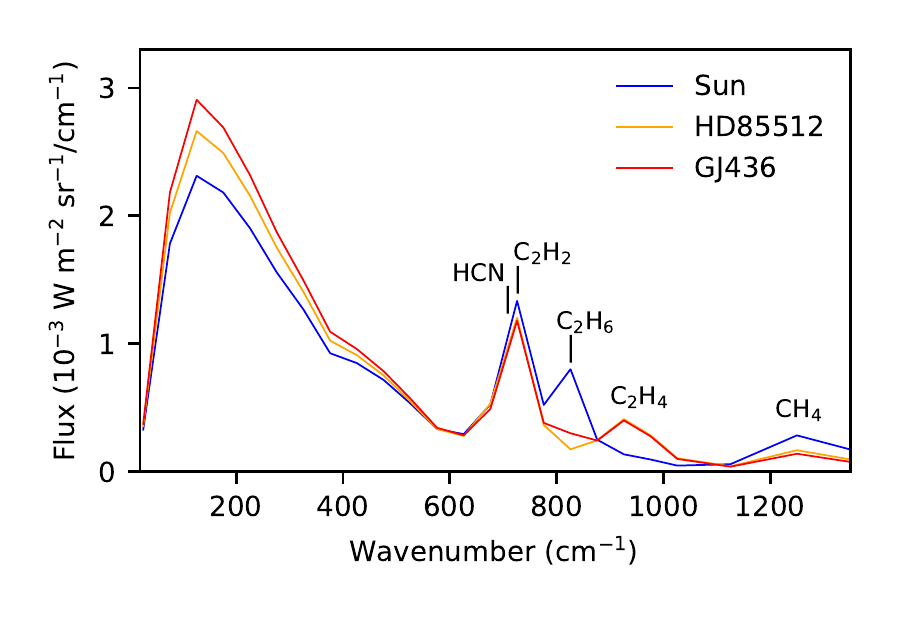}
   \caption{Global-mean thermal emission spectra. Spectra of planets around G (blue curve), K (orange curve), and M (red curve) stars, with spectral features due to radiatively active species labeled.}\label{emiss_spectra}
\end{figure}

Using the results from our TAM and KINETICS simulations, we produce globally-averaged planetary emission spectra for each of our three cases. These are shown in Fig.~\ref{emiss_spectra} as a function of wavenumber. In the case forced by solar radiation, we note good agreement with the thermal flux spectra of Titan's atmosphere derived using the Cassini Composite Infrared Spectrometer (CIRS) \citep{tomasko+2008b}, despite the relatively coarse resolution of our radiative transfer. 

All of our results show prominent spectral features due to the trace absorbers. The features due to HCN and $\mathrm{C_2H_2}$ occur with similar magnitudes in all three cases, whereas methane is apparent in all cases but with a more prominent feature in the case of the G dwarf host star. This difference results from the differences in temperature profiles of the three atmospheres (Fig.~\ref{temp_profiles}), since abundance profiles for methane in the pressure range captured by the GCM are virtually indistinguishable (Fig.~\ref{tracegasprofiles}). On the other hand, the ethane feature is obvious in the Solar case but difficult to make out and essentially nonexistent in the M and K dwarf cases, respectively. These latter simulations in turn feature a prominent $\mathrm{C_2H_4}$ peak that is not present in the G dwarf case. These spectral differences result directly from the abundance differences produced by the photochemistry (Fig.~\ref{tracegasprofiles}).

At the longwave end of the spectra, differences in total flux are also apparent between the simulations. The considerable emission between 10 and 600 cm$^{-1}$, which is due to molecular pairs (CIA), is higher for the planet irradiated by the M dwarf, with the K and G dwarf host star cases producing progressively lower fluxes. A similar overall trend occurs at short (near-IR) wavelengths, where methane absorption bands become increasingly prominent (not shown). These spectral signatures, in conjunction, thus illustrate various important features that could be used to diagnose Titan-like exoplanetary atmospheres around various stars.

We do not include geometric albedo spectra at short wavelengths in this study, because these would be almost entirely shaped by the presence of haze and therefore identical between the three cases, since we do not vary the haze structure. However, in analyzing the effect of enhanced haze production on Titan reflection spectra, \cite{checlair+2016} previously found that the slope from UV to visible wavelengths transitioned from a negatively-sloped Rayleigh-dominated regime, to a positively-sloped spectrum dominated by haze, with increasing haze production rates.  Given that our results tentatively suggest similar haze production for all three stellar host types, we would therefore expect that short wavelength spectra of such planets would closely resemble Titan's.

\pagebreak

\section{Discussion and Conclusions} \label{sec:conclusion}

We have presented asynchronously coupled simulations of the atmospheric circulation and photochemistry of Titan-like exoplanets to test their sensitivity to host stellar type. We explored the effects of three different host stellar spectra on the temperature structure, dynamics, and composition of the atmosphere of putative planets with otherwise identical insolation distributions and planetary properties. Despite the differences in the regions of absorption of stellar light---and consequent subtle differences in the temperature profiles, meridional circulation depths, and zonal wind structures---arising primarily from the interaction of haze with the varying insolation spectra, our simulations produced generally similar atmospheric structures in all cases. Differences in the chemical makeup of trace species in these Titan-like atmospheres are more complicated, but arise primarily from the differences in Lyman-$\alpha$ flux coming from the three different host stars. Acetylene and hydrogen cyanide profiles are very similar, while ethylene, ethane, and methane profiles have noticeable differences (in the latter case only at the lowest pressures). Intriguingly, our results tentatively suggest that haze formation---at least in the column integral---is similar in all cases. 

While our simulations explore the effects of host stellar type, we have not investigated the response to variations of other climate-relevant parameters. For terrestrial planets, the rotation rate, atmospheric mass, surface gravity, and planetary radius all act to influence the atmospheric circulation and equator-to-pole differences---and hence are important factors in determining atmospheric dynamics, chemistry, and habitability \citep{kaspi+showman2015}. Variations of these parameters will also affect the atmospheres of Titan-like exoplanets; however, we have opted to focus exclusively on host star type in order to robustly test the sensitivity of the atmosphere to incoming stellar radiation. The impact of these other parameters will be explored in a controlled way in future work, in order to make further progress in understanding the key physical mechanisms shaping exoplanetary atmospheres and their emergent spectra. In particular, we plan to explore how the circulation and consequent observable properties of similar cool planets depend on the planetary radius, as larger bodies are closer to characterization with facilities such as WFIRST, LUVOIR, HabEx and OST.  

We have simulated the photochemistry in our models using global-mean properties, and also ignored potential seasonal effects that would result if the planets' orbital eccentricities and obliquities were nonzero. On Titan, seasonally changing regional insolation and circulation patterns result in dramatic contrasts in the concentration of some species with latitude \citep{teanby+2013}, and in local condensation clouds that further contribute to the interaction of the atmosphere with radiation \citep[for example,][]{griffith+2006,anderson+2014}. Such complexities may be relevant for future work, but we consider our simplifications appropriate for our first-order exploration of Titan-like exoplanets.

The structure of Titan's present atmosphere depends on the presence of methane (and its photochemical products), and in this work we have considered only putative Titan-like exoplanetary atmospheres with similar methane content. Titan's atmospheric methane, however, has a finite lifetime of tens of millions of years \citep{yung+1984}, and there is evidence that the current carbon inventory is only approximately half a billion years old \citep{mandt+2012} whereas the nitrogen is essentially primordial \citep{niemann+2005}. It is therefore possible that Titan-like nitrogen atmospheres can exist for long periods largely devoid of methane, punctuated by episodic replenishment \citep{tobie+2006,tobie+2009}. ``Snowball'' Titan states would exhibit different atmospheric structure \citep{lorenz+1997b,charnay+2014}, as well as very different photochemistry \citep{wong+2015}, potentially with different dependences on host stellar spectra. Such hypothetical atmospheres, for which we have no observational constraints, are beyond the scope of the present study.

One of the most important source of uncertainty for our conclusions stems from our lack of knowledge about haze formation in general. The presence and structure of haze are critical for the atmospheric thermal structure, thus impacting the circulation. Our results suggest that overall haze production rates might be similar for a range of stellar spectra, implying that very Titan-like exoplanets could exist around a wide range of host stars. However, it should be noted that we rely on a parameterization of haze production using haze precursors to arrive at our estimated production rates, and that the actual chemical pathways leading to haze formation are as yet poorly understood \citep[see][for a review]{horst2017}. For example, both models \citep[e.g.,][]{lavvas+2013} and observations \citep{coates+2007,crary+2009,liang+2007,wahlund+2009,waite+2007} of Titan's atmosphere indicate that ion chemistry plays an important role in haze formation; our photochemical model lacks ion chemistry, and therefore our haze formation picture is incomplete. Laboratory experiments investigating the chemical precursors, pathways, and energy sources that lead to Titan-like hazes and their resulting composition \citep{cable+2012,horst+2013,horst+2017,imanaka+smith2010,trainer+2013,sciamma-obrien+2014} are critical to improving our understanding of the formation of exoplanetary hazes, including its dependence on external factors like stellar forcing. Thus, the combination of such ongoing laboratory work and modeling like ours---along with various observations, including short wavelength studies that might probe the existence of haze \citep{checlair+2016}---is required to make further progress in characterizing the diversity of hazy, cool, terrestrial exoplanets.

\acknowledgments

We thank Nikole Lewis for helpful discussions and detailed comments on the manuscript. We also thank Y. L. Yung, C. Li, R.-L. Shia, and S. Fan for discussions on the details of Titan photochemistry. This research was carried out in part at the Jet Propulsion Laboratory, California Institute of Technology, under a contract with the National Aeronautics and Space Administration.  P. Gao acknowledges support from an NAI Virtual Planetary Laboratory grant from the University of Washington to the Jet Propulsion Laboratory and California Institute of Technology under solicitation NNH12ZDA002C and Cooperative Agreement Number NNA13AA93A, the NASA Postdoctoral Program, and the 51 Pegasi b Fellowship funded by the Heising-Simons Foundation.

\end{document}